\documentclass[aps,pre,twocolumn]{revtex4}

\usepackage{amsmath}
\usepackage{amssymb}
\usepackage{amsbsy}
\usepackage{makeidx}
\usepackage{epsf}
\usepackage{graphicx}
\usepackage{amsfonts}
\usepackage{subfigure}

\newcommand{\sech}{\operatorname{sech}}
\newcommand{\csch}{\operatorname{csch}}

\newcommand\beq{\begin{equation}}
\newcommand\eeq{\end{equation}}
\usepackage{amsmath}
\usepackage{amssymb}
\newcommand{\ep}{\epsilon}
\newcommand{\lag}{\mathcal{L}}

\begin{document}
\title{Variational approximations to homoclinic snaking in continuous and discrete systems}
\author{P.C. Matthews}
\author{H. Susanto}
\affiliation{School of Mathematical Sciences, University of Nottingham, University Park, Nottingham, NG7 2RD, UK}


\begin{abstract}
Localised structures appear in a wide variety of systems,
arising from a pinning mechanism due to the presence of a small-scale
pattern or an imposed grid.  When there is a separation of
lengthscales, the width of the pinning region is exponentially small
and beyond the reach of standard asymptotic methods.
We show how this behaviour can be obtained using
a variational method, for two systems.
In the case of the quadratic-cubic Swift-Hohenberg equation,
this gives results that are in agreement with recent work
using exponential asymptotics. Secondly, the method is applied to a
discrete system with cubic-quintic nonlinearity, giving results
that  agree well with numerical simulations.
\end{abstract}

\maketitle

\section{Introduction}

This paper is concerned with the phenomenon of pinning of fronts
in nonlinear dynamical systems.
For systems that exhibit bistability of two uniform states,
a front connecting these two states will normally drift in one
direction, depending on which of the two states is preferred. At a
particular parameter value, known as the Maxwell point,
there is no preference between the two states and a stationary front exists.
However, in systems where there is an underlying structure, on a scale
that is typically small compared with the lengthscale of the front,
a front can become locked to this structure.
This mechanism allows a stationary front to exist over a range of
parameter values around the Maxwell point, known as the pinning
region. Placing two fronts back-to-back creates a localized state, and
the bifurcation diagram plotting the length of this localized solution
against the control parameter has a snaking structure, involving a
sequence of saddle-node bifurcations in near-perfect alignment. Since the
spatial structure of such a localized state departs from and then
returns to, a uniform state, the phenomenon has become known as
``homoclinic snaking''.
See \cite{knob08,dawe10} for a review of the subject and some open
questions.

There are two distinct scenarios in which homoclinic snaking appears.
In the first of these, the underlying structure is provided by pattern
formation. A physical system, for example convection in fluids
\cite{bens88,nepo94}, a vibrated granular material, buckling of a
solid cylinder \cite{hunt00}, optical systems \cite{firt07} or gas discharge
experiments \cite{will90,will91},  has an instability leading  to the formation
of a regular periodic pattern as a parameter is varied.
If this bifurcation is subcritical, there is bistability between the
uniform and the patterned states.  Near the onset of pattern
formation, a subcritical Ginzburg-Landau equation can be derived
\cite{bens88,budd05,burk07_2}, which has front-like solutions connecting the two states.
However, this equation does not capture the locking mechanism.
The localised states have been found numerically in many studies
\cite{saka96, burk06, beck09} that use the Swift-Hohenberg equation as a
simple model for pattern formation.

A second situation in which locked fronts appear is in discretized
forms of partial differential equations, where there is a locking
effect to the imposed lattice.  Examples include stationary solutions
of the discrete bistable nonlinear Schr\"odinger equation
\cite{carr06,chon09,chon11}, which leads to a subcritical Allen-Cahn
equation \cite{tayl10}, optical cavity solitons \cite{yuli08,yuli10},
and discrete systems with a weakly broken pitchfork bifurcation
\cite{cler11}.  This situation also arises whenever a bistable system
is solved numerically on a spatial grid.

An interesting and challenging question is to determine the width in
parameter space of the region in which stationary localized solutions
exist.
Numerical simulations indicate that this pinning region becomes very
small as the separation of the two length-scales increases
\cite{saka96, burk07_2}. In fact the region is exponentially
small, or beyond all orders. This is related to the fact that a conventional
multiple-scales asymptotic method cannot describe the locking effect,
since it regards to the two lengthscales as independent, while in fact
the locking mechanism involves an explicit interaction between the
short and long lengthscale.
The necessary exponential asymptotics calculations, involving
truncating a divergent asymptotic series at an optimal point
and obtaining an equation for the exponentially small remainder term,
have recently been carried out for the Swift-Hohenberg equation with
quadratic-cubic nonlinearities \cite{kozy06,chap09} and cubic-quintic
nonlinearities \cite{dean11}. However, these calculations are
extremely cumbersome and include an undetermined constant that must
either be obtained numerically either by fitting to numerical results
\cite{chap09}
or by approximately solving a complicated recurrence relation \cite{dean11}.

In this paper we use variational methods to obtain scaling laws for
the structure of the snaking region, building on the work in a
previous short paper that studied the cubic-quintic Swift-Hohenberg
equation \cite{susa11}. Of course, not all systems that exhibit
homoclinic snaking
have a variational structure, but most of those previously studied
do. The variational method is dependent on good initial ansatz,
but this is known in the cases studied here.
For pattern forming problems near onset, it is known that the solution
can be described by a slowly varying envelope function multiplied by
a sinusoidal wave. Furthermore, the form of the envelope function is
known from the standard asymptotic analysis, and can be obtained to
higher order if required.  Thus there is no guesswork involved in the
method. Calculating the integrals in the Lagrangian automatically
leads to exponentially small terms, and from these we can obtain the
phase of the locked states that is inaccessible to the usual
asymptotic expansion. Note that variational methods have been used
before to study localized states on lattices \cite{carr06,chon11}, but
not in the slowly-varying regime that is studied here. Also in the
continuous case, variational methods have been used \cite{wade00}, but
not in the snaking regime.

Here, we consider two equations, namely the quadratic-cubic Swift--Hohenberg equation and the discrete cubic-quintic Schr\"odinger equation, also known as the spatially discrete Allen--Cahn equation, representing continuous and discrete systems, respectively. The first equation is discussed in Section \ref{sec:qc}. The results are then compared with numerical results obtained by continuation in Section \ref{sec:num}. The second equation is studied in Section \ref{sec:dac}. The scaling calculated analytically using the variational method is then compared with computational results, where good agreement is obtained. Conclusions are in Section IV.

\section{The quadratic-cubic Swift-Hohenberg equation}
\label{sec:qc}

The quadratic-cubic Swift-Hohenberg equation is given by
\begin{equation}
\partial_tu=ru-\left( 1+\partial_x^2\right)^2u+b_2u^2-b_3u^3.
\label{gov1}
\end{equation}
This equation represents a simple model for pattern forming systems
that do not have a symmetry under sign reversal of the dependent
variable $u$, and has been very widely used to illustrate homoclinic
snaking \cite{beck09,budd05,burk06,kozy06}.
The Lagrangian for (\ref{gov1}) is
\begin{equation}
\mathcal{L}=\int_{-\infty}^\infty \left( \frac{u_{xx}^2}{2}-{u_x^2}+(1-r)\frac{u^2}2
-\frac{b_2}{3}u^3+\frac{b_3}4u^4 \right)dx.
\label{lag1}
\end{equation}
It can easily be shown that $\mathcal{L}$ decreases with time, so
stable stationary states of (\ref{gov1}) correspond to minima
of (\ref{lag1}).
It will be assumed that $b_3>0$; in this case $u$ can be rescaled to
set $b_3=1$.
In (\ref{gov1}), the bifurcation at $r=0$ is
subcritical (allowing localised patterns) if
$b_2 > \sqrt{27 b_3/38}$.

\begin{figure}
\includegraphics[width=8.8cm]{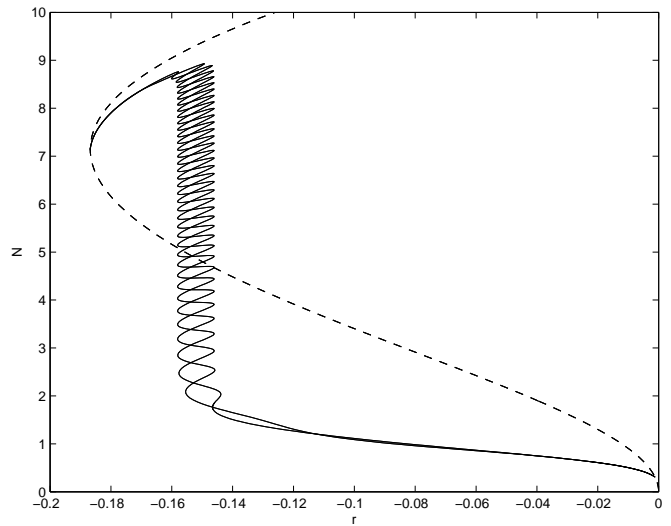}
\caption{Bifurcation diagram of (\ref{gov1}) for $b_2=1.5$,
  $b_3=1$. The dashed line is the periodic solution. The two solid
  lines are the localized snaking solutions.}
\label{fig:qcsnake}
\end{figure}

\begin{figure}
\includegraphics[width=8.8cm]{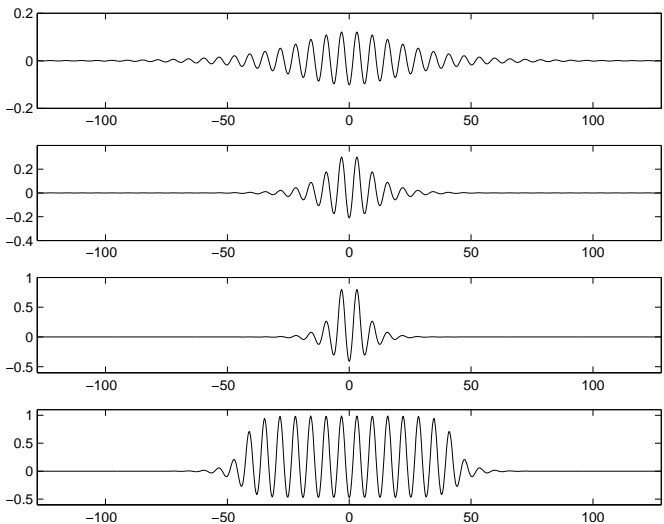}
\caption{Localized solutions on one branch of
  Fig.~\ref{fig:qcsnake}. From top to bottom, $r=-0.01$, $r=-0.05$,
  $r=-0.146$, $r=-0.158$ ($N=5.15$).}
\label{fig:loc}
\end{figure}

Figure~\ref{fig:qcsnake} shows the bifurcation diagram for
(\ref{gov1}) obtained by a numerical continuation method
for $b_2=1.5$,
$b_3=1$, with periodic boundary conditions in a domain of length
$l=82\pi$. The norm $N$ plotted is defined by
\begin{equation}
N^2 = \int_{-l/2}^{l/2} u^2 \, dx .
\label{norm0}
\end{equation}
The periodic solution (dashed line) bifurcates subcritically and
becomes stable at a saddle-node bifurcation at $r\approx -0.186$.
Two branches of localized solutions bifurcate from the periodic state
at a very small value of $r$ and form an intertwined snaking pattern
near the Maxwell point at  $r\approx -0.151$.
On one of these branches, the maximum of the envelope function
coincides with a minimum of the periodic pattern,
and on the other it coincides with a maximum, so the two states have a
phase difference of $\pi$.
Solutions along the former branch are shown in
Figure~\ref{fig:loc}. The first figure, for $r=-0.01$, shows a slow
spatial modulation that can be represented by an envelope in the form
of a sech function (see Sec.~\ref{sec:sech}). For $r=-0.05$ the
amplitude modulation occurs over a shorter lengthscale but can still be
represented by the sech function. The third graph shows $u(x)$ at the
first saddle-node bifurcation,  $r=-0.146$, and the final graph is at
a saddle-node bifurcation higher up the graph where $r=-0.158$ and
$N=5.15$. At this stage the solution resembles two  fronts connecting the
periodic solution to the state $u=0$ (see Sec.~\ref{sec:max}).

\subsection{Analysis of solutions near $r=0$}
\label{sec:sech}

The analysis near $r=0$ can be performed in a similar way to that in \cite{susa11}. Motivated by the results from multiple scale expansions \cite{yang97,wade00,burk06}, we take the ansatz
\begin{equation}
u=A\sech(Bx)\cos(kx+\varphi)+C\sech^2(Bx), \label{sech23}
\end{equation}
from which we obtain that the norm (\ref{norm0}) is given by
\begin{widetext}
\begin{eqnarray}
N^2&=&\frac1{3B}\left(3A^2+4C^2\right)+\frac{3A\pi}{B^3}\left(C(B^2+k^2)\cos(\varphi)\sech\left(\frac{k\pi}{2B}\right)
+ABk\cos(2\varphi)\csch\left(\frac{k\pi}{B}\right)\right).
\end{eqnarray}

Substituting the ansatz (\ref{sech23}) into the Lagrangian \eqref{lag1} yields the effective Lagrangian
\begin{eqnarray}
L_{eff}&=&\frac{2C^2}{315B}\left(36b_3C^2-56b_2C-168B^2+240B^4+105-105r\right)\nonumber\\
&&-\frac{A^2}{30B}\left(-24b_3C^2+20b_2C-7B^4-30k^2B^2+10B^2-15k^4+30k^2+15(r-1)\right)+\frac{b_3}{8B}A^4\nonumber\\
&&+
{\frac {\pi  A k^2}{720 B^7}} \left( -60 b_2 C^2 k^2
 B^2+45\, A^2 b_3\, C\, B^2 k^2+B^4\left(240 C k^4 -180 A^2 b_2 +720 C(1-k^2) \right) +2 b_3 C^3 k^4\right)\nonumber\\
&& \times\cos(\varphi)\left(e^{-\frac{k\pi}{2B}}+\mathcal{O}\left(e^{-\frac{k\pi}{B}}\right)\right).
\label{lagr2}
\end{eqnarray}
\end{widetext}
It is immediately clear from (\ref{lagr2}) that the phase $\varphi$ is
determined by
exponentially small terms, since the parameter $B$ is expected to be
small, with $1/B$ representing the lengthscale of the modulation of
the pattern. Furthermore, (\ref{lagr2}) shows that steady states
(extrema of $L_{eff}$) exist if $\varphi$ is a multiple of $\pi$
\cite{yang97}. There are two distinct states,
$\varphi=0$ and $\varphi=\pi$, both corresponding to even solutions
but with the maximum of the envelope function $\sech(Bx)$ coinciding
with a maximum or minimum of the wave $\cos(kx)$, as shown by the
numerical continuation method in the previous section.
Note that this result does not depend on our choice of ansatz. For any
slowly varying, even modulation function $f(Bx)$, the integrals in the
Lagrangian involve powers of $f(Bx)$  and its derivatives multiplied
by $\cos(kx)\cos(\varphi)$, yielding an exponentially small quantity
multiplied by $\cos(\varphi)$, and hence stationary states with
$\varphi = 0,\pi$.
In the symmetric cubic-quintic case \cite{susa11}, only even powers appear in
the integrals, leading to integrals only involving $\cos(2kx)\cos(2\varphi)$,
and hence stationary solutions with
$\varphi = 0,\pi/2,\pi,3\pi/2$.

Applying the Euler-Lagrange formulation to the effective Lagrangian
\begin{equation}
\frac{\partial}{\partial\alpha}\mathcal{L}_{eff}=0,
\label{euler}
\end{equation}
gives us a system of nonlinear equations for $\alpha$, with $\alpha=A,\,B,\,C,\,k,\,\varphi$, that make (\ref{sech23}) an approximate solution of (\ref{gov1}).

Neglecting the exponentially small terms in the effective Lagrangian (\ref{lagr2}), the Euler-Lagrange equations can be solved perturbatively about $r=0$ to yield
\begin{eqnarray}
A&=&\sqrt{\frac8{4b_2^2-3b_3}}\sqrt{-r}+\mathcal{O}((-r)^{3/2}),\label{Acq}\\
B&=&\frac12\sqrt{-r}+\mathcal{O}(-r),\\
C&=&-\frac{4b_2}{4b_2^2-3b_3}r+\mathcal{O}((-r)^{3/2}),\\
k&=&\sqrt{1-B^2}=1+\frac{r}8.\label{kcq}
\end{eqnarray}

It is important to note that the presence of $C\neq0$ plays an
important role in the calculations above, unlike the case of
the cubic-quintic equation \cite{susa11} where it was possible to
include only the first term in (\ref{sech23}). Taking $C=0$ would
result in the leading
order expression of $A$ independent of $b_2$, which is incorrect.
This occurs because when $C=0$ the cubic term in the Lagrangian,
which corresponds to the crucial symmetry-breaking quadratic term in
(\ref{gov1}), does not contribute to the Lagrangian integral except through an
exponentially small term.
Note that $C=b_2 A^2/2$, a result that is also easily be obtained
at second order in the asymptotic analysis of the problem.
A similar Lagrangian method was used by Wadee and Bassom
\cite{wade00}, using more terms in the ansatz (\ref{sech23}) but with
$\varphi=0$.
The results above for $A$ and $C$ are almost the same as those
obtained using multiple
scale expansions \cite{wade00,burk06}. A slight difference arises as
we have for simplicity omitted a second-order term in
$\sech^2(Bx)\cos(2kx+2\varphi)$; including this term would give exact
agreement between the Lagrangian method and the asymptotic analysis.
According to (\ref{Acq}) the transition from a sub- to supercritical
bifurcation occurs at $b_2^2/b_3=3/4$, which is very close to the true
value of 27/38 \cite{kozy06}.

The equation (\ref{kcq}) indicates a slight decrease in the
wavenumber, corresponding to a slight increase in the wavelength of
the patterns. As noted by Wadee and Bassom \cite{wade00}, this is
simply a consequence of the linear dispersion relation; (\ref{kcq}) can
be obtained directly from the linear terms in (\ref{gov1}) when
$r<0$. Hence it is not surprising that the same wavenumber correction
was found in the cubic-quintic case \cite{susa11}.

When neglecting the exponentially small terms, the phase-shift $\varphi$ at this order is arbitrary, which also agrees with the multiple scales result. However, taking into account the equation $\partial_\varphi \mathcal{L}_{eff}=0$, in which all the terms are exponentially small, $\varphi$ has to be a multiple of $\pi$, as mentioned above.

The ansatz (\ref{sech23}) is only appropriate for small values of $|r|$,
away from the Maxwell point (see Figs.~\ref{fig:qcsnake} and \ref{fig:loc}).
In the snaking region near the Maxwell point, the envelope of the
localized states resembles two connected fronts.
The following section introduces a suitable ansatz for this regime.

\subsection{Analysis of solutions near the Maxwell point}
\label{sec:max}

For values of  $b_2$ in the neighbourhood of
\begin{equation}
b_{20}=\sqrt{\frac{27}{38}b_3}, 
\label{subsup}
\end{equation}
the bifurcation is only slightly subcritical and the Maxwell point is
within reach of weakly nonlinear analysis.
In this case the quadratic-cubic Swift-Hohenberg equation (\ref{gov1})
has a front solution, which is approximately given by \cite{kozy06}
\begin{equation}
u=A_M\frac{\cos{\left(x-\frac1{2\sqrt{734}}\ln\left(1+e^{a_Mx}\right)\right)}}{\sqrt{1+e^{-a_Mx}}},
\label{fs}
\end{equation}
with
\begin{eqnarray}
A_M&=&\sqrt{\frac{38\sqrt{-r_M}}{\sqrt{734}b_3}},\quad a_M=\sqrt{-r_M},
\label{AM1}\\
r_M&=&-\frac{6859}{17616 b_3}(b_2-b_{20})^2 .\label{rM1}
\end{eqnarray}

In analytically estimating the width of the snaking region in the
equation using variational approximations, we will employ a front
solution similar to (\ref{fs}). Considering this solution, one can
note that the oscillation wave number $k$ of the front changes in
space. In the limits $x\to\pm\infty$, $k\to1$ and
$(1-a_M(2\sqrt{734})^{-1})\approx1$. Therefore,
we will fix $k=1$ to simplify the analysis.

It turns out that as in the case of the $\sech$ solutions considered
in the previous section, if only the leading order front solution
(\ref{fs}) is used as the ansatz,  the leading terms in the
Lagrangian do not depend on $b_2$, giving qualitatively incorrect
results. We therefore adopt an ansatz of the form
\begin{equation}
u=\frac{A\cos(x+\phi)}{\sqrt{1+e^{B(|x|-L)}}}
+ \frac{b_2 A^2}{2(1+e^{B(|x|-L)})}.\label{front23}\end{equation}
The second term here arises at second order in the  weakly nonlinear
expansion \cite{wade00}, included for a similar reason as discussed in the previous section.
This term is responsible for the upward displacement of the periodic
pattern that is apparent in Fig.~\ref{fig:loc}.
To simplify the calculation we have set the value of the coefficient of
this term in advance rather than leaving it as a free parameter.
The asymptotic analysis includes another second-order term,
proportional to $\cos 2(x+\varphi)$ \cite{wade00}, which is omitted partly in the
interests of simplicity and partly because it is smaller by a factor
of 9 than the term we have included.

Evaluation of the Lagrangian integral is considerably more complicated
than in the cubic-quintic case \cite{susa11}. This is partly because of the
two-term ansatz and partly because the dominant terms come from odd
powers of the square root function, requiring integrals that
cannot simply be computed using residues. These integrals can be
evaluated in terms of multiplications of gamma functions of a complex
argument, which in turn can be expanded using Stirling's formula when
$B$ is small, giving the anticipated exponentially small terms.
There are a large number of these exponentially small terms,
and the calculation is complicated by the fact that terms involving
higher powers of $u$, which one might expect to give a smaller
contribution, in fact all give exponentially small terms of the same
order. 

To illustrate some of the steps in the calculation, consider the term
$\int_{-\infty}^\infty u^2 dx$, which is the defined norm (\ref{norm0}) and one of the terms in the Lagrangian density (\ref{lag1}). Using the ansatz, this can be written as
\begin{eqnarray}
&&\int_{-\infty}^\infty u^2 dx = \int_{0}^\infty \frac{A^2\left(1+b_2^2A^2/2+e^{B (x-L)}\right)}{\left(1+e^{B (x-L)}\right)^2}\nonumber\\
&&+\frac{2b_2A^3\cos(\phi)\cos(x)}{\left(1+e^{B (x-L)}\right)^{3/2}}
+\frac{A^2\cos(2 \phi) \cos(2 x)}{1+e^{B (-L+x)}}\, dx,
\end{eqnarray}
after using the double-angle formula and the fact that the
envelope function is even. The first term, that is independent of $\phi$, in the integral can be
evaluated directly and gives the answer $A^2 L\left(1+A^2 b_2^2/2 \right)-A^4 b_2^2/(2 B) +\mathcal{O}(e^{-BL})$.
The integral of the third term can be found by using contour integration, similarly to that used in \cite{susa11}. Yet, we are not interested in the explicit expression of it, as it contributes to higher harmonic corrections in $\phi$. The second term is of our interest, giving a leading exponentially small contribution of order $\mathcal{O}(e^{-\frac{\pi}{B}})$. Unfortunately, the integral cannot be immediately evaluated by using the residue method as mentioned above. By using Mathematica, one will obtain that
\begin{widetext}
\begin{eqnarray}
&&\int_{0}^\infty \frac{\cos(x)\,dx}{\left(1+e^{B (x-L)}\right)^{3/2}}  = \frac{i e^{-i L} \Gamma\left(\frac{3B+2i}{2B}\right) \Gamma\left(\frac{-i+B}{B}\right)}{\sqrt{\pi }}-\frac{i e^{i L} \Gamma\left(\frac{3B-2i}{2B}\right)\Gamma\left(\frac{i+B}{B}\right)}{\sqrt{\pi }}\nonumber\\
&&+\frac{e^{B L}}{B} {\left[\frac{2 e^{B L} \sqrt{1+e^{-B L}}}{1+e^{B L}}-{}_2F_1\left(\frac{i-B}{B},\frac{1}{2},\frac{i}{B},-e^{-B L}\right)-{}_2F_1\left(-\frac{i+B}{B},\frac{1}{2},-\frac{i}{B},-e^{-B L}\right)\right]}, \label{u2}
\end{eqnarray}
\end{widetext}
where $\Gamma$ and ${}_2F_1$ are respectively the Gamma and the hypergeometric function.

The hypergeometric function can be written in a series form as
\[
{}_2F_1(a,b,c,z)=\sum_{n=0}^\infty\frac{(a)_n(b)_n}{(c)_n}\frac{z^n}{n!},
\]
where $(\bullet)_n=\bullet(\bullet+1)\dots(\bullet+n-1)$ is the Pochhammer symbol. The series is convergent in our case, i.e.\ $0<B\ll1$ and $BL\gg1$, such that up to $\mathcal{O}(e^{-2BL})$
\begin{eqnarray*}
&&{}_2F_1\left(\frac{i-B}{B},\frac{1}{2},\frac{i}{B},-e^{-B L}\right)=1-\frac12 e^{- B L} \left(1+iB\right),\\
&&{}_2F_1\left(-\frac{i+B}{B},\frac{1}{2},\frac{-i}{B},-e^{-B L}\right)=1-\frac12 e^{- B L} \left(1-iB\right).
\end{eqnarray*}

As for the Gamma function, it can be approximated by the Stirling's formula
\[
\Gamma(z)=\sqrt{\frac{2\pi}{z}}\left(\frac{z}{e}\right)^z\left(1+\mathcal{O}(\frac1z)\right),
\]
when $z$ is large enough in absolute value. Then, the first two terms on the right hand side of (\ref{u2}) can be approximated by
\begin{eqnarray}
\frac{e^{-\frac{\pi}{B}}}{2 B^{3/2}} \sqrt{\frac{\pi }{2}} \big((-8+3 B) \cos(L)+(8+3 B) \sin(L)\big).
\end{eqnarray}

Combining the approximations above, one will obtain that
\begin{eqnarray}
&&\int_{-\infty}^\infty u^2 dx = A^2 L\left(1+\frac{1}{2} A^2 b_2^2 \right)-\frac{A^4 b_2^2}{2 B}\nonumber\\
&&+\frac{A^3 b_2 }{B^{3/2}} \exp\left(-\frac{\pi }{B}\right)\sqrt{\frac{\pi }{2}} \cos(\phi)\nonumber\\
 &&\times\bigg((-8+3B) \cos(L)+(8+3 B) \sin(L)\bigg).
 \label{U1}
\end{eqnarray}
The formula (\ref{U1}) indicates that in the snaking region, solutions with a larger norm correspond to longer plateaus.

Performing the same calculations to all the terms in the Lagrangian
(\ref{lag1})  yields the effective Lagrangian
\begin{widetext}
\begin{eqnarray}
\mathcal{L}_{eff}&=&\frac{A^2}{1920B}(15B^4+480B^2-160A^2b_2^2B^2+16A^2b_2^2B^4+480A^2b_2^2+480A^2rb_2^2\nonumber\\
&&+240A^4b_2^4-360A^2b_3-1080A^4b_3b_2^2-110A^6b_3b_2^4)\nonumber\\
&&+\frac{LA^2}{96}(18A^2b_3-24A^2b_2^2-48r-24A^2rb_2^2-8A^4b_2^4+36A^4b_3b_2^2+3A^6b_3b_2^4)\nonumber\\
&&+\cos(\varphi){\frac {{A}^{3}\sqrt {2}\sqrt {\pi }b_2}{420}} {e^{{{-\pi }/{B}}}}{B}^{-7/2}\left(K_+\cos(L)+K_-\sin(L)\right)\nonumber\\
&& +\mathcal{O}\left(e^{\frac{-2\pi}{B}}(\sin(2L),\cos(2L)),e^{-BL}\right),
\label{efl3}
\end{eqnarray}
\end{widetext}
where
\begin{eqnarray}
K_\pm&=&
\pm56\,{A}^{4}b_2^{2}{b_3}+280\,{A}^{2}b_2^{2}B-420\,{A
}^{2}{b_3}\,B\nonumber\\
&&\pm156\,{B}^{2}\pm840\,r{B}^{2}-763\,{B}^{3}+70\,r{B}^{3}.
\label{Kpm0}
\end{eqnarray}


Consider first the terms that are not exponentially small, that is,
the terms that do not involve the phase $\varphi$.
Bearing in mind the anticipated scaling from (\ref{AM1}),
$A=\mathcal{O}(r^{1/4})$, the
dominant terms in the bracket multiplied by $L$ are the first two
terms, so $\mathcal{L}_{eff}$ only has a minimum over $L$ if at
leading order, $b_2^2={3b_3/4}$. This is very close to the
asymptotic result (\ref{subsup}), and would be exactly the same if we
had included the $\cos 2(x+\varphi)$ term mentioned above.
We therefore set
\begin{equation}
b_2=\sqrt{3b_3/4}+\Delta,\quad |\Delta|\ll1,
\end{equation}
and expect that $\Delta$ and $B$ are $\mathcal{O}(r^{1/2})$,
in which case the leading terms in the Lagrangian are
\begin{eqnarray}
\mathcal{L}_{eff}^{lead}&=&\frac{A^2}{128B}(32B^2+32\sqrt{3b_3}A^2\Delta-45A^4b_3^2)
\nonumber\\
&& + \frac{L A^2}{96}(45A^4b_3^2/2-48r-24\sqrt{3b_3} A^2\Delta ).
\end{eqnarray}
Finding and solving the Euler-Lagrange equations by differentiating
with respect to $L$, $A$ and $B$ gives the values
\begin{equation}
A^2 = \frac{8\sqrt{3} \Delta}{15 b_3^{3/2}}, \quad
B = \frac{\sqrt{10}\Delta}{5\sqrt{b_3}}
\end{equation}
and yields the Maxwell point
\begin{equation}
r_M = - \frac{2\Delta^2}{5 b_3} = - B^2,
\end{equation}
which is within $3\%$ of the asymptotic value given in (\ref{rM1}).
Note that the length $L$ is not determined by these equations.
Substituting these values of $A$, $B$ and $r_M$ into the
leading terms of the Lagrangian  we find that the minimum is
\begin{equation}
Min(\mathcal{L}_{eff}^{lead}) = \frac{4\sqrt{30}\Delta^2}{75b_3^2}.
\end{equation}
Thus, although the Maxwell point can be determined by the condition
that $\mathcal{L}$ is zero for both the zero state and the periodic
state, the value of $\mathcal{L}$ is not zero for the localized state,
due to the presence of the fronts. In fact,  $\mathcal{L}$ is positive
and of the same order as $r$.

Considering now the exponentially small term involving $\varphi$,
it is clear from $\partial_\varphi\mathcal{L}_{eff}=0$ that there are
two distinct branches of stationary snaking solutions, with
$\varphi=0$ and $\varphi=\pi$, corresponding to the centre
of the localised pattern being a local maximum or minimum
respectively.
However, another possible way to satisfy
$\partial_\varphi\mathcal{L}_{eff}=0$
is that the coefficient multiplying the $\cos\varphi$ term vanishes, i.e.
\begin{equation}
K_+\cos(L) + K_-\sin(L) = 0 .
\end{equation}
Substituting the above values of $A$ and $B$, and taking the leading
terms in $K_\pm$, solutions of this type occur for values of $L$
given by
\begin{equation}
\tan L = K_+/K_-,
\label{L0}
\end{equation}
where $K_\pm$ is given in (\ref{Kpm0}), which to the leading order is
\begin{equation}
K_\pm={\frac {8\sqrt [4]{30}\sqrt {\pi }}{7875\,b_3}}\,{ { \left(\pm 307\,\sqrt {3}-210\,\sqrt {10} \right) }{{
}}}.
\label{Kpm}
\end{equation}
Hence, for small $B$ these states exist if $L=-0.110219... + m \pi$ for integer values of $m$.
These solutions have been referred to as ``bridges'' \cite{wade02}
or ``ladders''  \cite{burk07} and were identified in the exponential
asymptotics of (\ref{gov1}) by Yang and Akylas \cite{yang97} and
Chapman and  Kozyreff \cite{chap09}.
The ladder states (not shown in Fig.~\ref{fig:qcsnake}) have no
reflection symmetry  and connect the two snaking branches, bifurcating
from them near the saddle-node bifurcations.

To find the snaking range, we set $\sin(\varphi)=0$ and
$r=r_M+\delta r$, and the Lagrangian simplifies to
\begin{eqnarray}
\mathcal{L}_{eff} &=& \frac{4\sqrt{30}\Delta^2}{75b_3^2}-\frac{\delta rA^2L}{2}\nonumber\\
 &&+e^{-\frac{\pi}B}\left(K_+\cos(L)+K_-\sin(L)\right),
 \label{efl23s}
\end{eqnarray}
with $K_\pm$ given in (\ref{Kpm}). This can be minimised over $L$ to give
the exponentially small value of $\delta r$ for stationary solutions,
\begin{equation}
{\delta r}=\frac{2e^{-\frac{\pi}B}}{A^2}\left(K_-\cos(L)-K_+\sin(L)\right).
\end{equation}
The maximum value of $\delta r$ (half the width of the snaking region)
is then, using the leading-order approximations
for $b_2$, $A$ and $B$ given above,
\begin{equation}
\delta r_m = {\frac {2e^{-\pi/B}}{525{\Delta}}}\,{{\sqrt {241249\pi b_3}
\sqrt [4]{120}}}.
\label{drm23}
\end{equation}
Note that the dependence of the snaking width on the small parameter
$\Delta$ is very similar to that for the small parameter
$b_3$ in the cubic-quintic case \cite{susa11}. The dependence in (\ref{drm23})
is the same as that obtained by Kozyreff and
Chapman \cite{kozy06} using the methods of exponential asymptotics,
although the comparison is not immediate since they regard $b_2$ as
the control parameter rather than $r$. Note that the numerical
coefficient was determined by fitting in \cite{kozy06}.

\begin{figure}
(a)
\includegraphics[width=8.cm]{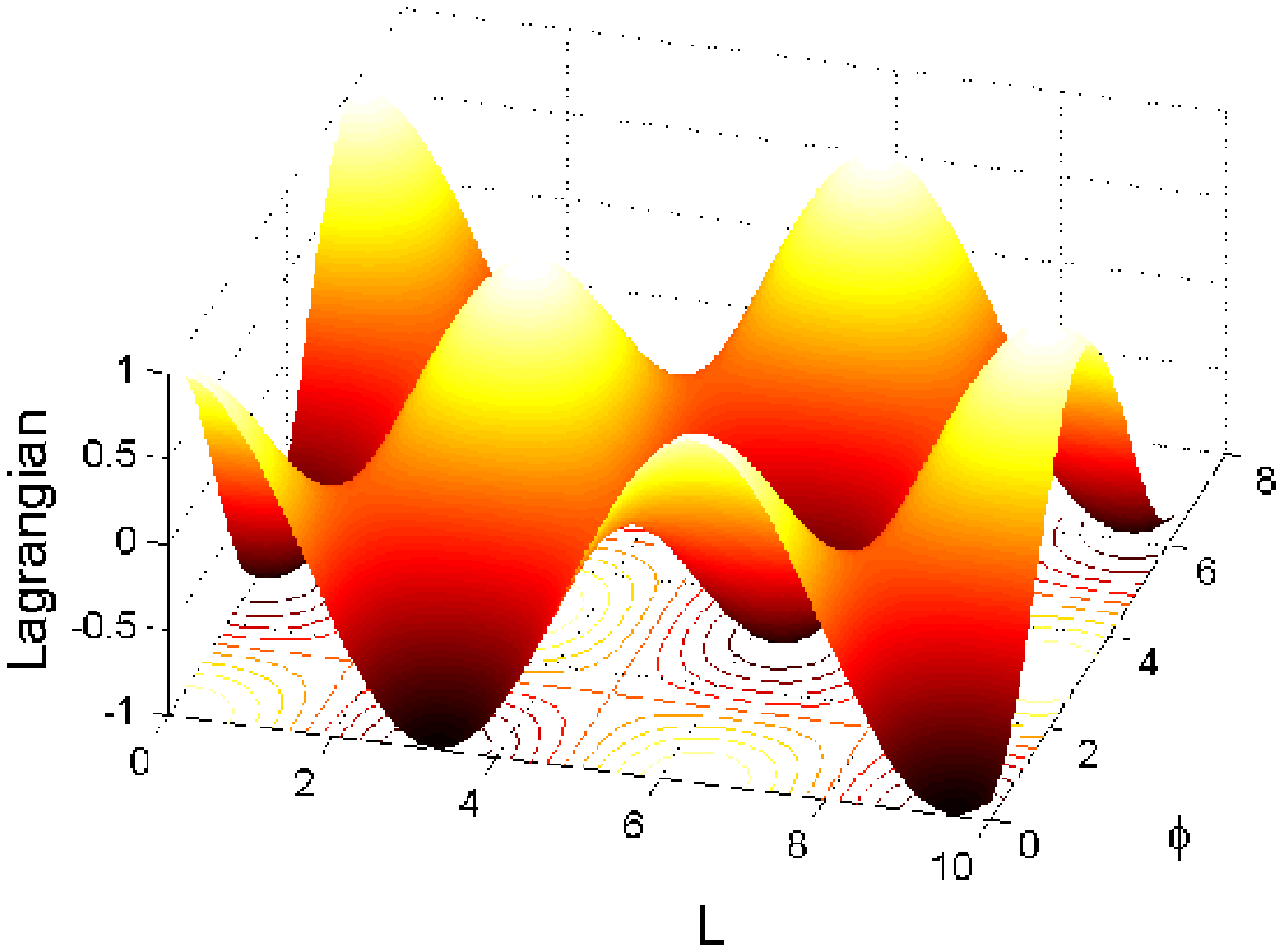}
(b)
\includegraphics[width=8.cm]{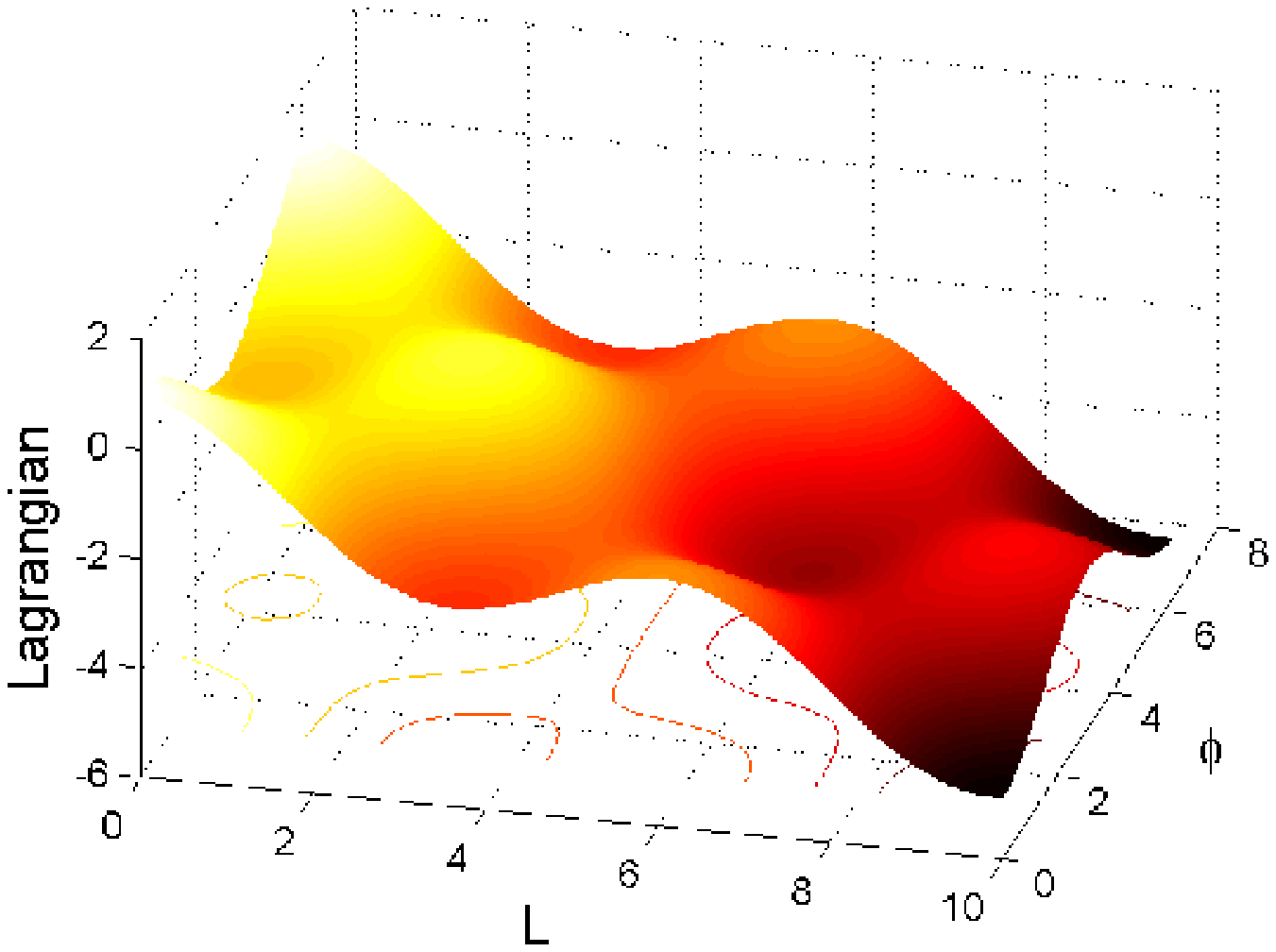}
\caption{Surface plots of the simplified Lagrangian (\ref{eq:le}), (a)
with $\delta r = 0$ and (b) with  $\delta r = 0.4$.}
\label{fig:le}
\end{figure}

A useful feature of the Lagrangian method is that we can examine the
stability of steady states since we know that stable equilibria are
local minima of the Lagrangian.
Considering the exponentially small terms, with $A$ and $B$ fixed and
$r=r_M+\delta r$, the dependence of  $\mathcal{L}$ on $\delta r$, $L$ and
$\varphi$ can be represented by the function
\begin{equation}
\mathcal{L}_e(\delta r,L,\varphi)
= - \delta r L + \cos\varphi \cos(L-L_0),
\label{eq:le}
\end{equation}
where the phase shift $L_0$ satisfies (\ref{L0}) and for convenience we have set the constants to 1.
Figure~\ref{fig:le}(a) shows a surface plot of $\mathcal{L}_e$ for
$\delta r = 0$. The solutions on the snaking branches, with
$\varphi=0, \pi$, $L-L_0=m\pi$ are either maxima or minima, so are either stable or
unstable with two positive eigenvalues. The ladder solutions at
$\varphi=\pi/2, 3\pi/2$, $L-L_0=\pi/2+m\pi$ are saddle points and are
therefore always unstable.
The surface plot for $\delta r=0.4$ is shown in Fig.~\ref{fig:le}(b).
The symmetric snaking solutions remain at $\varphi=0, \pi$ but are
shifted in $L$, while the ladder states are shifted in $\varphi$ but
remain at the same values of $L$.

\subsection{Numerical results}
\label{sec:num}

\begin{figure*}[tbhp]
\centering
\subfigure[]{\includegraphics[width=7cm,clip=]{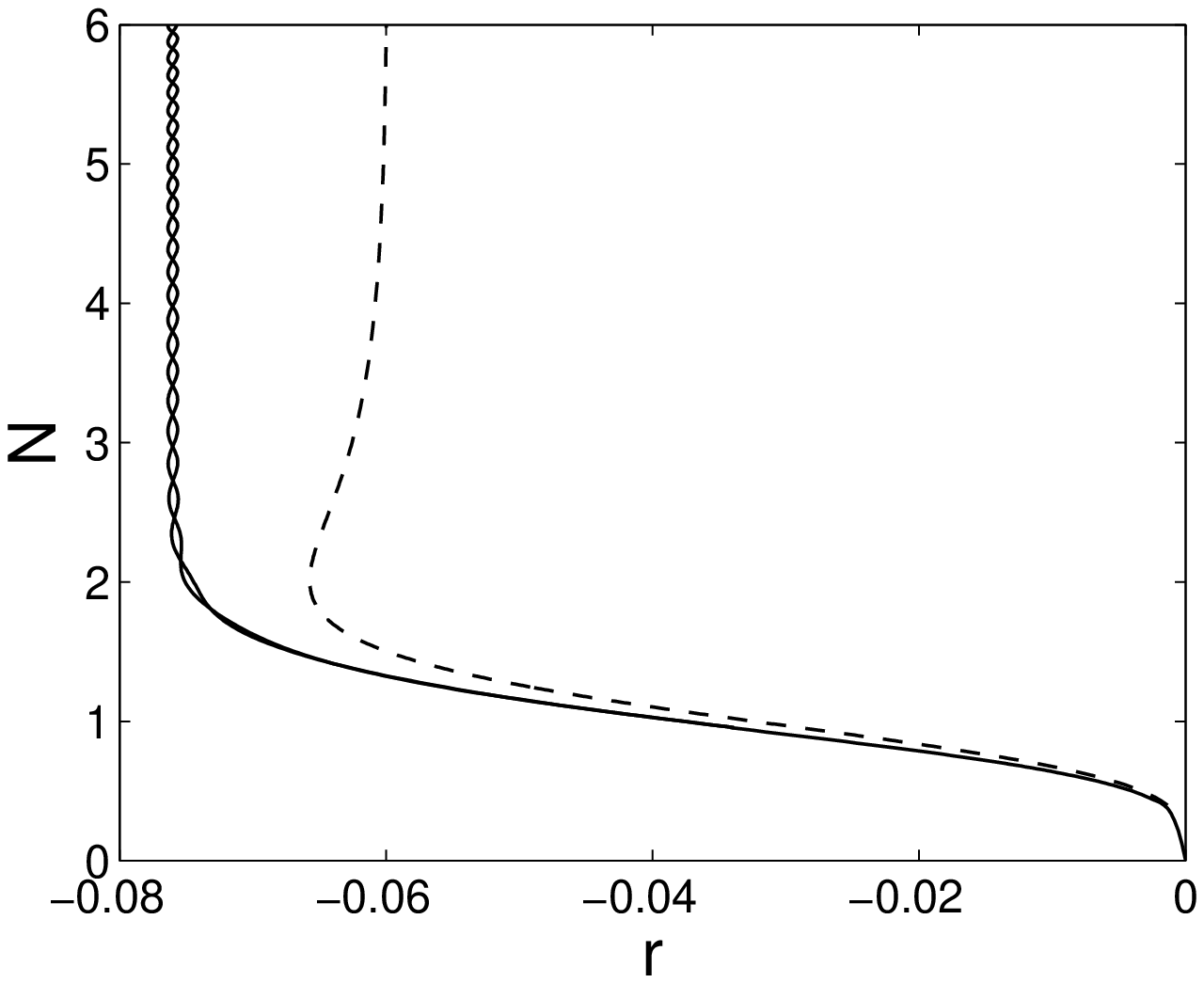}}
\subfigure[]{\includegraphics[width=7cm,clip=]{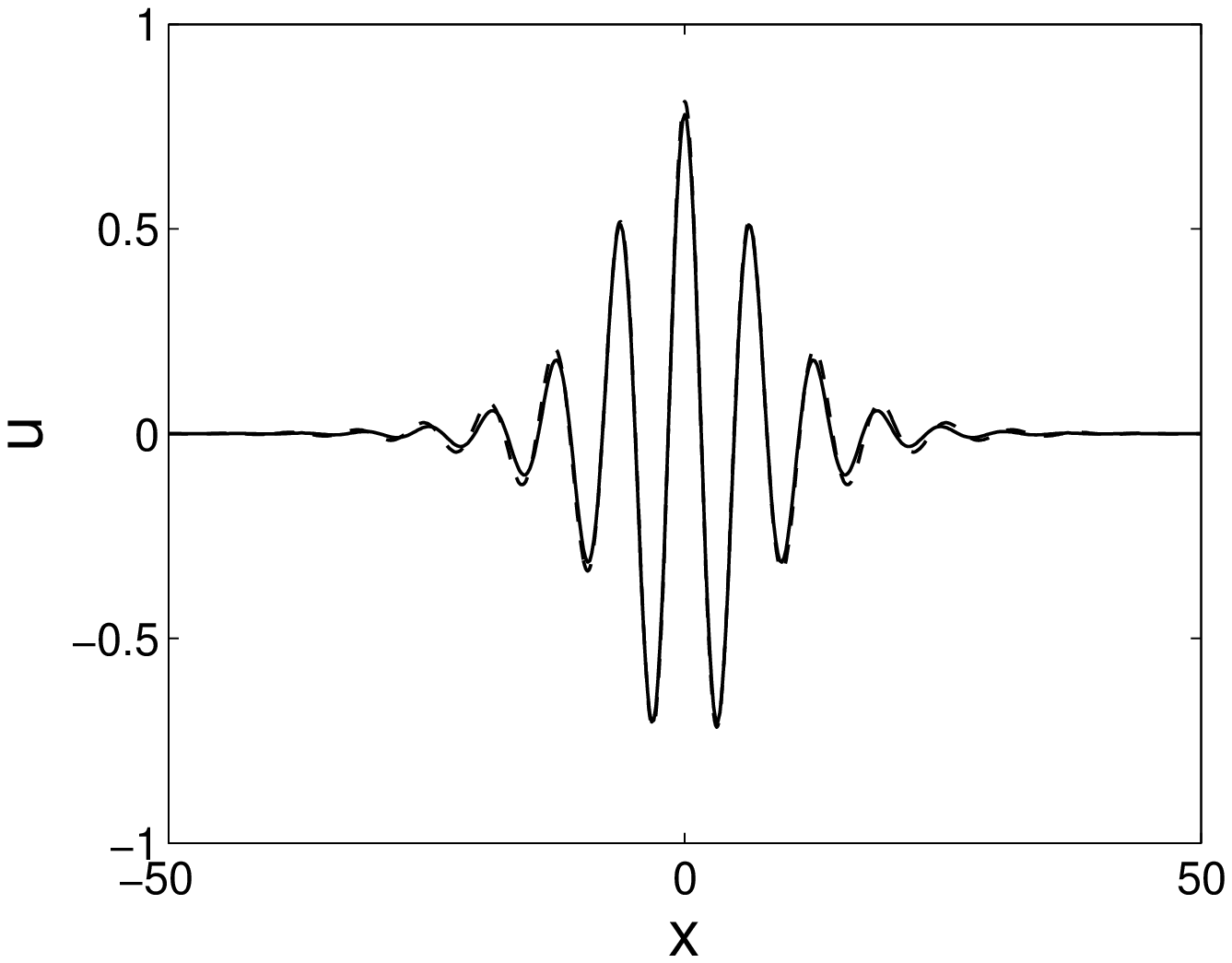}}\\
\subfigure[]{\includegraphics[width=7cm,clip=]{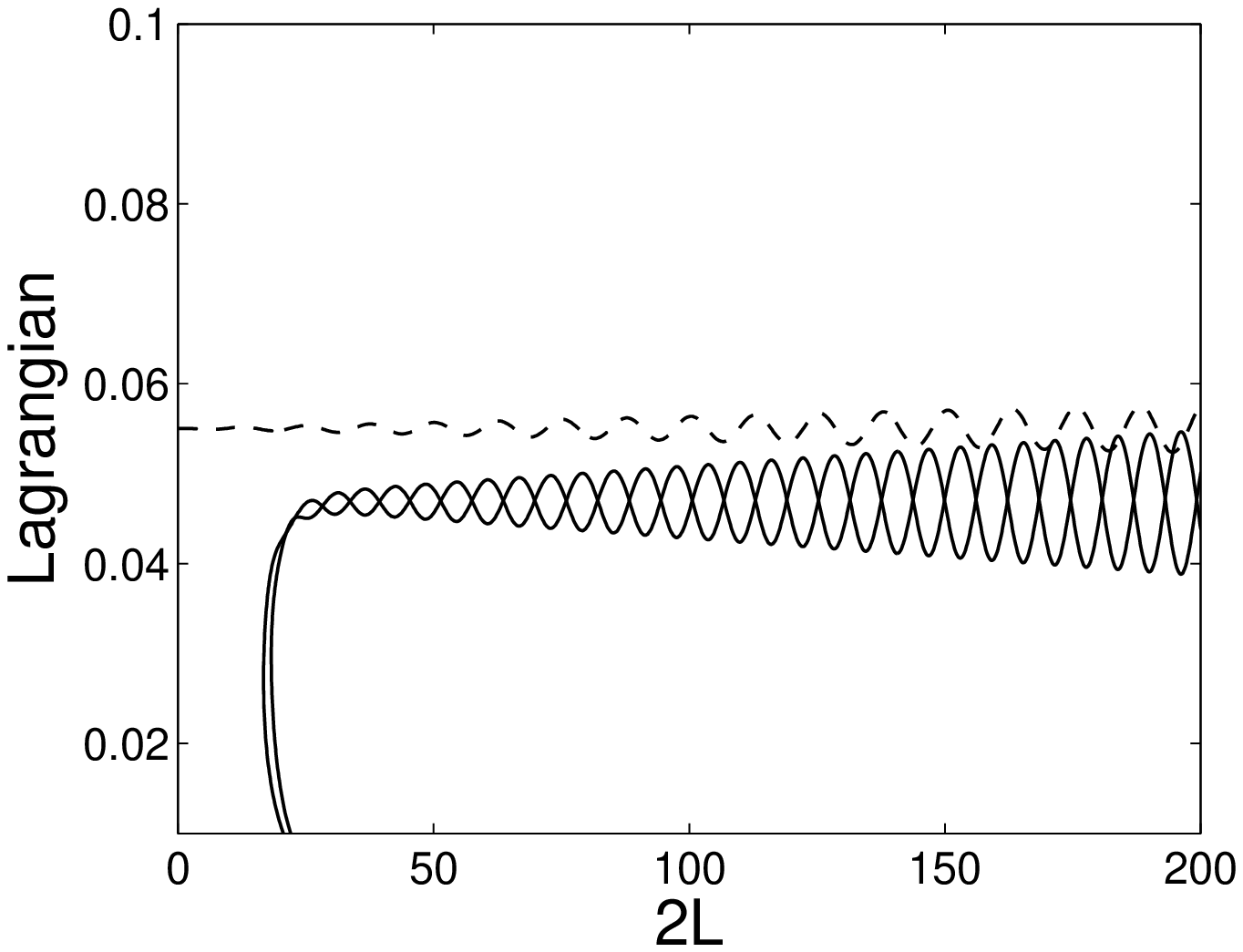}}
\subfigure[]{\includegraphics[width=7cm,clip=]{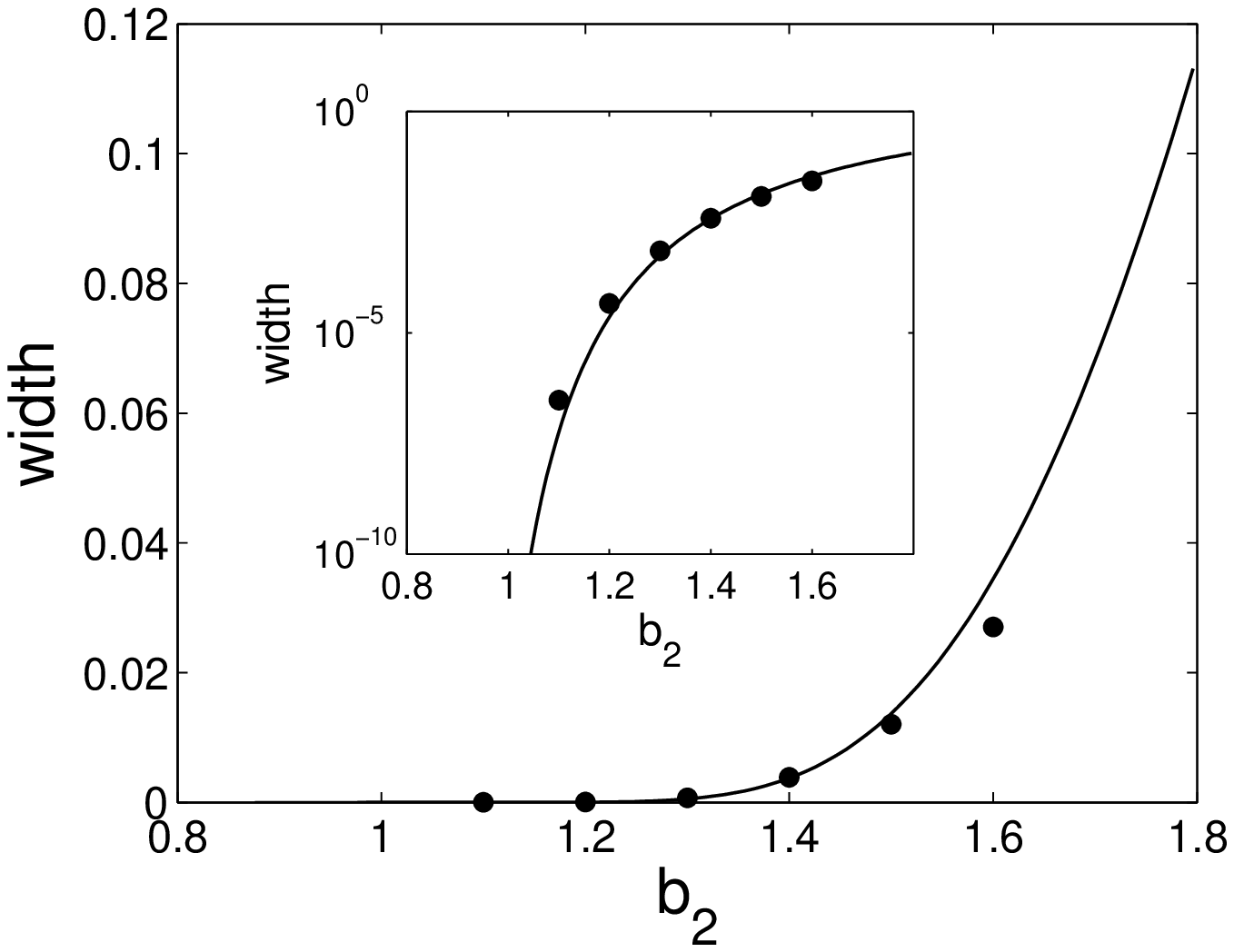}}
\caption{(a) The bifurcation diagram obtained numerically and our approximation obtained from solving (\ref{euler}) with the ansatz (\ref{sech23}) for $b_2=1.3$. (b) A numerically obtained localized solution and its approximation from the variational formulation for  $\varphi=0$ and $r=-0.06$. (c) The numerically computed Lagrangian (\ref{lag1}) as a function of the length of the plateau $2L$, corresponding to panel (a). 
The dashed line is our approximation calculated from the effective Lagrangian (\ref{efl23s}). (d) The width of the snaking region as a function of $b_2$. Filled circles are numerical and solid lines are analytical, i.e.\ $2\delta r_m$ (\ref{drm23}). The inset shows the same comparison in a log scale. In all the figures, $b_3=1$.}\label{figQC}
\end{figure*}

In this section we compare the above results from the variational
method with numerical solutions of
the quadratic-cubic Swift-Hohenberg equation
(\ref{gov1}).  We have solved the equation numerically for localized
states, using a pseudo-arclength continuation method with periodic
boundary conditions, implemented with a Fourier spectral
discretization. Plotted in the top left panel of Fig.\
\ref{figQC} is the bifurcation diagram showing two branches of
localized solutions for $b_2=1.3$ and $b_3=1$. In the same panel, shown
in dashed lines are our analytical results obtained from
(\ref{euler}) using the sech ansatz, where one
can see that the variational calculation approximates the numerics
very well for relatively small $|r|$. As $r$ decreases further and enters
the snaking region, the approximation deviates from the numerics. In
the top right panel, we plot the profile of a localized solution at
$r=-0.06$ and its approximation.


Considering Eqs.\ (\ref{Acq})--(\ref{kcq}) one can conclude that for
larger values of the parameter $b_2$, the parameter $|r|$ can be
larger while  keeping the amplitude $A$ (\ref{Acq})  small. This
implies that the sech ansatz can have a longer validity region for
large $b_2$. 
It is important to note that the ansatz might be able to quantitatively capture the first snaking bifurcation for a larger value of $b_2$.

In panel (c) of the same figure, we plot the Lagrangian (\ref{lag1}) corresponding to the bifurcation diagram (a) as a function of the length of the solution's plateau $2L$, which is calculated numerically as
\begin{equation*}
\displaystyle 2L\approx\frac{8 \int_{-l/2}^{l/2} u^2\,dx}{(u_M-u_m)^2+2(u_M+u_m)^2},
\end{equation*}
where $u_M=\max\{u\}$ and $u_m=\min{\{u\}}$. The integration is a definite integration over the computational domain.
Plotted in the same panel is our effective Lagrangian (\ref{efl23s}). There is good agreement for the average numerical value of the Lagrangian and the qualitative nature of the oscillations, but the amplitude of the oscillations appears to be underestimated in the variational approximation. Note that the amplitude of the oscillations in $\mathcal{L}$
increases with $L$ since from (\ref{efl23s}) with $\delta r \propto \cos(L)$, there are oscillations of the form  $L  \cos(L)$.

Finally, in panel (d) we show the width of the snaking region as a function of $b_2$ numerically and analytically, where one can see that our approximation is in good agreement.

The numerical method can also be used to check the stability of the
steady state solutions (since the Jacobian is already computed as part
of the Newton-Raphson iteration). This confirms that the snaking
states are in general either stable, or unstable with two small
positive eigenvalues. One eigenvalue changes sign at the saddle-node
bifurcation and another at the nearby bifurcation to the ladder
states.  The ladder states are found to be unstable with one positive
eigenvalue, consistent with the fact that they are saddle points of the
Lagrangian.

\section{Snaking in discrete systems}
\label{sec:dac}

To illustrate the wide-ranging applicability of the variational method
in describing the snaking behaviour, we analyse in this section an
analogous problem in discrete systems.  This turns out to be
remarkably similar to the continuous case.
Consider the system of ordinary differential  equations
\beq
\frac{du_n}{dt} = C (u_{n+1} - 2 u_n + u_{n-1} ) + \mu u_n + 2 u_n^3 - u_n^5,
\label{disc}
\eeq
with the condition that $u_n \to 0$ as $n\to \pm \infty$.
We assume that $C >0$ and look for stationary solutions of (\ref{disc}).
This system has been considered in \cite{carr06,chon09,chon11,tayl10},
and very similar systems have been  studied in
\cite{yuli08,cler11}.
The system (\ref{disc}) is interpreted as a discrete form of the
nonlinear Schr\"odinger equation in  \cite{carr06,chon09},
and a discrete subcritical Allen--Cahn equation in \cite{tayl10}.
These previous studies have used numerical continuation to
obtain bifurcation diagrams that show a very
similar snaking structure to that seen in the continuous
Swift-Hohenberg equation (\ref{gov1}).
Note however that (\ref{disc}) is {\em not} a discrete form of the
Swift-Hohenberg equation. Rather it is a discrete form of the
subcritical Ginzburg--Landau equation that describes the slowly varying
envelope function for (\ref{gov1}).
Thus the snaking behaviour seen in  (\ref{disc}) results from a
locking mechanism to the discrete lattice, with the lattice points
providing the analogue of the small-scale pattern in the
Swift-Hohenberg equation.  In view of this, computational
investigation of  the snaking in (\ref{disc}) is much easier than in
(\ref{gov1}), since there is no small-scale structure to resolve.

The grid-locking behaviour of fronts in (\ref{disc}) can be seen as an
indication of numerical error in the finite-difference approximation
of the  bistable partial differential equation
\beq
\frac{\partial u}{\partial t} = \frac{\partial^2 u}{\partial X^2} + \mu u + 2 u^3 - u^5.
\label{accont}
\eeq
When (\ref{accont}) is discretized with second-order finite
differences  with mesh spacing $\ep$, the result is  (\ref{disc})
with $C=\ep^{-2}$.
Front-like solutions to (\ref{accont}) are travelling waves, for all
values of $\mu$ except exactly at the Maxwell point. Thus the snaking
branch of (\ref{disc}) indicates a range of values of $\mu$ where the
numerical approximation incorrectly finds a stationary solution.
It is therefore of interest to determine the width of this region
for small values of $\ep$ (large values of $C$).
An alternative scaling, similar to that in
Sec.~\ref{sec:qc}, is to set $C=1$ and replace the 2 in (\ref{disc})
by a small parameter $b_2$; however we will use the form (\ref{disc})
to facilitate comparison with the previous work cited above.

The Lagrangian for (\ref{disc}) is
\beq
\lag=\sum_{n=-\infty}^\infty \frac C 2 (u_{n+1}-u_n)^2 - \frac \mu 2
u_n^2 -\frac 1 2 u_n^4 + \frac 1 6 u_n^6.
\label{ldisc}
\eeq
Note that a variational method has been used in \cite{carr06,chon11} to study
the case of small $C$, but here the focus will be on large $C$.
The uniform solutions of  (\ref{disc}) are given by $u_n=0$ and
\beq
u_n^2 = 1 \pm \sqrt{1+\mu},
\label{usoldisc}
\eeq
so there is a saddle-node bifurcation at $\mu=-1$ and there is
bi-stability of both the zero state and the larger non-zero solution in
(\ref{usoldisc}) for $-1 < \mu < 0$.
By finding the Lagrangian of this non-zero state it is straightforward
to show that the Maxwell point is at $\mu = -3/4$ and
so at this point the stable non-zero state is $u_n = \sqrt{3/2}$.

For large $C$ we expect the solution to (\ref{disc}) to be close to
that of the continuous system (\ref{accont}).
At the Maxwell point  there is an exact stationary front-like solution of
(\ref{accont}), $u(X)$, given by
\beq
u^2 = \frac {A^2}{1+e^{\alpha X}}, \mbox{ where } A^2=3/2, \ \alpha=\sqrt{3}.
\eeq
In the discretized system, $u_n= u(\ep n)$, and we wish to consider a
localized state linking two fronts.  The ansatz analogous to that used in
the continuous case (\ref{front23}) is then
\beq
u_n^2 = \frac {A^2}{1+e^{\alpha\ep(|n-\phi|-L)}}. \label{discanz}
\eeq
Here, the length of the front (measured in terms of the number of mesh
points) is $2L$ and $\phi$ is a phase variable, which is arbitrary at
this stage but will be determined by exponentially small terms.
If $\phi=0$ then $u_j = u_{-j}$ and the localized state is symmetrical and
`site-centred'. Similarly, a `bond-centred' symmetrical state, with
the property $u_j = u_{1-j}$, can be obtained by setting $\phi=1/2$.
Note that it is not necessary to include higher order terms in
(\ref{discanz}) as in (\ref{front23}); this is because (\ref{disc})
only has nonlinear terms with odd powers and therefore is more
analogous to the cubic-quintic Swift-Hohenberg equation analysed with
the Lagrangian approach in \cite{susa11}.

Consider now the sum $\sum_{n=-\infty}^\infty u_n^2$
that appears in $\lag$.  This can be evaluated by writing the sum as
an integral of a function multiplied by a sum of $\delta$ functions,
and then writing the sum of $\delta$ functions as a Fourier series:
\begin{eqnarray*}
\sum_{n=-\infty}^\infty u_n^2 &=&
    \sum_{n=-\infty}^\infty \frac {A^2}{1+e^{\alpha\ep(|n-\phi|-L)}} \\
&=& \int_{-\infty}^\infty  \frac {A^2}{1+e^{\alpha\ep(|x-\phi|-L)}}
\sum_{n=-\infty}^\infty \delta(x-n) \, dx \\
&=& \int_{-\infty}^\infty  \frac {A^2}{1+e^{\alpha\ep(|x-\phi|-L)}}
\sum_{k=-\infty}^\infty e^{2i k\pi x} \, dx .
\end{eqnarray*}
Note that $x$ here is not the same as $X$ in (\ref{accont}), rather it
is a continuous form of the variable $n$.  Of course the Fourier
series representation of the  $\delta$ function is not uniformly convergent,
but the integral converges very rapidly -- in fact exponentially.
From the term $k=0$ we obtain an order one contribution
\beq
\int_{-\infty}^\infty  \frac {A^2}{1+e^{\alpha\ep(|x-\phi|-L)}} \, dx
= 2 L A^2 + O(e^{-\alpha\ep L}).
\label{u2disc1}
\eeq
The terms with $k=\pm 1$ give a contribution
\beq
\int_{-\infty}^\infty \frac{ A^2 (e^{2i\pi x}+e^{-2i\pi x})}{1+e^{\alpha\ep(|x-\phi|-L)}} \, dx
\label{u2discep}
\eeq
which is exponentially small in $\ep$, and the terms arising from
larger values of $k$ are smaller by an  exponentially small factor.
Note that the integrals that are needed here, for the discrete case,
are essentially exactly the same integrals that were needed for the
continuous case in Sec.~\ref{sec:qc} and \cite{susa11}.

The integral (\ref{u2discep}) can be found using contour integration.
For the term in $e^{2i\pi x}$ the contour is closed in the upper half
plane and is dominated by the poles at $x=\phi\pm L +
i\pi/(\alpha\ep)$.  The sum of the residues at these two poles is
\begin{equation}
-2 i \sin(2\pi L) e^{2i\pi\phi}\frac {A^2}{\alpha \ep} e^{\frac{-2\pi^2}{\alpha\ep}}.
\end{equation}
The value of (\ref{u2discep}) is found by multiplying this by $2 i\pi$
and adding the complex conjugate to account for the $e^{-2i\pi x}$
term.
Hence the required sum is
\begin{widetext}
\beq
\sum_{n=-\infty}^\infty u_n^2 = 2 L A^2 + 8 \pi \sin(2\pi
L)\cos(2\pi\phi) \frac {A^2}{\alpha \ep} e^{\frac{-2\pi^2}{\alpha\ep}}
+ O(e^{\frac{-4\pi^2}{\alpha\ep}}, e^{-\alpha\ep L} ).
\label{sumu2disc}
\eeq
As in the continuous snaking case, we suppose that the exponential
terms responsible for the grid locking dominate those from the
interaction between the two fronts. This requires $L \gg \ep^{-2}$.

In a similar way it can be shown that
\beq
\sum_{n=-\infty}^\infty u_n^4 = 2 L A^4 - \frac {2 A^4}{\alpha\ep}
 - \frac {8\pi A^4}{\alpha^2\ep^2} \cos(2\pi\phi)
(2\pi\cos(2\pi L)-\alpha\ep\sin(2\pi L))e^{\frac{-2\pi^2}{\alpha\ep}}
\label{sumu4disc}
\eeq
and
\beq
\sum_{n=-\infty}^\infty u_n^6 = 2 L A^6 - \frac {3 A^6}{\alpha\ep}
 - \frac {8\pi A^6}{\alpha^3\ep^3} \cos(2\pi\phi)
(3\pi\alpha\ep\cos(2\pi L)+(2\pi^2-\alpha^2\ep^2)\sin(2\pi L))e^{\frac{-2\pi^2}{\alpha\ep}},
\label{sumu6disc}
\eeq
where again terms of order $e^{\frac{-4\pi^2}{\alpha\ep}}$  and
$e^{-\alpha\ep L}$ have been dropped. Note that for small $\ep$, the
dominant exponentially small term comes from (\ref{sumu6disc}).

The remaining term required for the Lagrangian is
\beq
\sum_{n=-\infty}^\infty  (u_{n+1}-u_n)^2
 = \sum_{n=-\infty}^\infty A^2 \left(\frac 1{1+e^{\alpha\ep(|n+1-\phi|-L)}}
 - \frac 1{1+e^{\alpha\ep(|n-\phi|-L)}}\right)^2 .
\label{discdiff}
\eeq
As with the other terms, this can be written as an integral,
\[
\int_{-\infty}^\infty A^2 \left(\frac 1{1+e^{\alpha\ep(|x+1-\phi|-L)}}
 - \frac 1{1+e^{\alpha\ep(|x-\phi|-L)}}\right)^2
(1+e^{2i\pi x} + e^{-2i\pi x} + \ldots) \, dx ,
\]
\end{widetext}
leading again to a part that is independent of $\phi$ and an
exponentially small $\phi$-dependent term.
For the part which does not depend on $\phi$, we can write the
integrand as $(f(X+\ep)-f(X))^2$, where $f(X)=A/\sqrt{1+e^{\alpha X}}$,
$X=\ep x$, and use Taylor expansion to find the dominant contribution to
(\ref{discdiff}) in the form
\[
\frac 1 4 A^2\alpha\ep -\frac 1 {384} A^2\alpha^3\ep^3 +O(\ep^5) .
\]
If the same method is applied for the exponentially small terms, it
turns out that each term in the Taylor expansion yields a contribution
that is of the same order after the exponentially small integral has
been evaluated.
Each term gives a contribution proportional to
\[
\frac{A^2}{\alpha\ep}\cos(2\pi\phi)\sin(2\pi L)e^{\frac{-2\pi^2}{\alpha\ep}},
\]
with a different numerical constant.
Thus we know the scaling but not the magnitude of this term.
Since (\ref{discdiff}) is multiplied by $C=\ep^{-2}$ in (\ref{ldisc}),
this term is of the same order as the $u^6$ term in (\ref{ldisc}).

Using the above results, the leading terms in the Lagrangian
(\ref{ldisc}) are
\beq
\lag_{lead}  = \frac {A^2\alpha}{8\ep} +\frac {A^4}{\alpha\ep}
 -\frac {A^6}{2\alpha\ep} + L\left(-\mu A^2-A^4+\frac{A^6}{3}\right).
\label{disclaglead}
\eeq
By making this expression stationary with respect to $A$, $L$ and
$\alpha$ we recover the results given earlier, $A^2=3/2$,
$\alpha=\sqrt{3}$, $\mu=-3/4$.

To find the width of the snaking region, we set $\mu=-3/4 +\Delta\mu$,
and introduce the largest of the exponentially small terms, arising
from the $u^6$ term and the difference term in (\ref{ldisc}), to obtain
\beq
\lag = \frac{3\sqrt3}{8\ep} -\frac 3 2 L \Delta\mu
-\frac {(8\pi^3+D) A^6}{3\alpha^3\ep^3}\cos(2\pi\phi)\sin(2\pi
L)e^{\frac{-2\pi^2}{\alpha\ep}},
\label{disclagsmall}
\eeq
where $D$ is an unknown constant representing the contribution from
(\ref{discdiff}) where only the scaling is known.

By making (\ref{disclagsmall}) stationary with respect to $\phi$
it follows that snaking branches branches exist with $\phi=0$ or
$\phi=1/2$, corresponding to the `site-centred' and `bond-centred'
states found numerically. For these snaking branches,
differentiating with respect to $L$ shows that
\[
\Delta\mu \propto \frac {1} {\ep^3}\cos(2\pi L) e^{\frac{-2\pi^2}{\alpha\ep}}
\]
so the scaling for the width of the snaking region in (\ref{disc}) for
large $C$ is
\beq
\Delta\mu \propto  {C^{3/2}} e^{-2\pi^2\sqrt{C/3}}.
\label{discwidth}
\eeq
Furthermore, differentiating (\ref{disclagsmall}) with respect to
$\phi$ predicts that the ladder states,
with $\sin(2\pi\phi)\ne 0$, occur
for integer values of $L$.

\begin{figure}
\includegraphics[width=8.6cm]{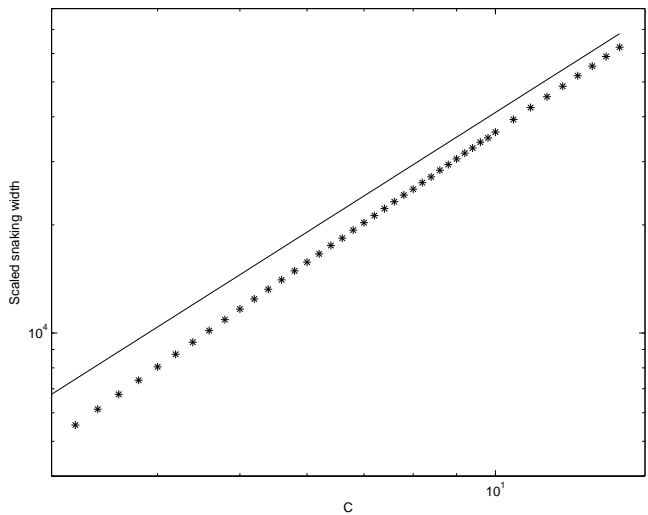}
\caption{Log-log plot showing numerically obtained values of $\Delta\mu
  e^{2\pi^2\sqrt{C/3}}$ (points) together with the $3/2$ scaling law
expected according to  (\ref{discwidth})  (line).}
\label{fig:discwidth}
\end{figure}

To check the scaling predicted by (\ref{discwidth}), snaking curves
for the equation were obtained numerically using a continuation software method.  With 200 grid points it was possible to obtain snaking bifurcations up to a value $C=14$ (at which point the snaking width is of the order of $10^{-14}$).
Figure~\ref{fig:discwidth} shows the numerically obtained snaking width multiplied by the predicted exponential factor
$e^{2\pi^2\sqrt{C/3}}$. This shows a clear power-law behaviour, with a power very close to the value $3/2$ predicted by (\ref{discwidth}).

\section{Conclusion}

In this paper we have used the variational approximation to study the
snaking behaviour of localised patterns in the quadratic-cubic
Swift--Hohenberg equation and the discrete bistable Allen--Cahn
equation.  With a simple ansatz, inspired by asymptotic analysis,
the exponentially small terms responsible for the snaking appear
in the Lagrangian. This enables the branches of snaking solutions to
be found, along with the asymmetric `ladder' states that link these
branches.

These solutions cannot be found by a conventional multiple-scales
method, since they involve a locking mechanism between the long and
short scales, but are accessible through  exponential asymptotics
\cite{kozy06,chap09}.
The Lagrangian approach provides a useful complement to the exponential
asymptotics method.  Both methods give the same scaling for the
relationship between the width of the snaking region and the small
parameter of the system.

We have shown that a close similarity exist between the pinning
phenomena in continuous and discrete systems. This arises partly
through the fact that the continuous limit of the discrete system
considered is exactly the same as the Ginzburg--Landau equation
describing spatial modulation of the pattern in the continuous case,
and partly because the sums in the Lagrangian for the discrete case
can be converted into integrals very similar to those that appear in
the continuous problem.

The variational method has a number of advantageous features, in
addition to the fact that the Lagrangian integral immediately
generates the necessary exponentially small terms.  Based on the
minimisation of the Lagrangian, it is easy to distinguish between
stable equilibria and unstable ones.  Furthermore, although we have
concentrated on the case of small parameters, for the purposes of
comparison with asymptotic methods, the method is not restricted to
this regime.  In future work it may be possible to apply the method to
cases where no parameters are small, and make useful comparisons with
numerical or experimental results. An additional challenge will be to
extend the results to two-dimensional systems
\cite{chon09,beck09,mcca10}.


\end{document}